\documentstyle[mprocl]{article}

\def\NPB{{\em Nucl. Phys.} B}
\def\PLB{{\em Phys. Lett.}  B}
\def\PRL{\em Phys. Rev. Lett.}
\def\PRD{{\em Phys. Rev.} D}
\def\ZPC{{\em Z. Phys.} C}
\def\be{\begin{equation}}
\def\ee{\end{equation}}
\def\bea{\begin{eqnarray}}
\def\eea{\end{eqnarray}}
\def\beq{\begin{equation}}
\def\eeq{\end{equation}}

\def\bea{\arraycolsep .1em \begin{eqnarray}}
\def\eea{\end{eqnarray}}
\def\Tr{{\rm Tr}}

\let\Ga=\Gamma

\def\eq#1{(\ref{#1})}

\def\s0#1#2{\mbox{\small{$ \frac{#1}{#2} $}}}
\def\0#1#2{\frac{#1}{#2}}

\begin{document}

\thispagestyle{empty}

\title{Mind The Gap}

\author{Daniel F. Litim}

\address{Theory Division, CERN, CH -- 1211 Geneva 23}

\maketitle\abstracts{ 
We discuss an optimisation criterion for the exact renormalisation group based on the inverse effective propagator, which displays a gap. We show that a simple extremisation of the gap stabilises the flow, leading to better convergence of {\it approximate} solutions towards the physical theory. This improves the reliability of truncations, most relevant for any high precision computation. These ideas are closely linked to the removal of a spurious scheme dependence and a minimum sensitivity condition. The issue of predictive power and a link to the Polchinski RG are discussed as well. We illustrate our findings by computing critical exponents for the Ising universality class.}

\vspace*{-8cm}
\begin{flushright}
{\normalsize 
CERN-TH-2001-13 
}
\end{flushright}
\vspace*{7cm}

\section{Motivation}

The exact renormalisation group (ERG) is an important method for studying non-perturbative problems in quantum field theory.~\cite{Polchinski,CW,Morris:1994qb} The particular strength of this formalism is its flexibility, allowing for systematic approximations without being tied to the small coupling region. The extension to gauge theories makes it a promising tool for strong interactions.~\cite{gaugefields} For theories as complex as QCD, the application of the ERG -- as of any other method -- is bound to certain approximations. It is known that the integrated full flow approaches the full quantum effective action, {independent} of the choice for the regulator. In turn, the solution to a {truncated} flow typically depends on the regulator in pretty much the same way as approximate computations in perturbative QCD depend on unphysical parameters.~\cite{Ball:1995ji,Litim96,Freire:2000sx}

The origin of this spurious scheme dependence is understood as follows. The ERG flow is induced by an infrared regulator, which is a highly non-local function of momenta coupling to all operators of the effective action. Therefore, a change of the regulator function modifies the effective interactions amongst all operators. While this has no effect for the full solution, it matters for truncated ones:  they are scheme dependent due to the absence of some neglected operators. Turning this observation around, it should be possible to identify optimised regulators which lead to a faster convergence of expansions, such that higher order contributions remain small and the physical information is almost exclusively contained in a few leading terms. This optimisation is of interest both conceptually and practically, and of great help for any high precision computation of physical observables.

In the present contribution, we discuss an optimisation criterion introduced in ref.~\cite{Litim:2000ci}. It is based on the infrared regulator, which, by definition, induces a gap for the effective inverse propagator. We explain why the extremisation of the gap leads to an improved convergence of the flow. The intimate link to the issue of spurious scheme dependence of approximate solutions is established. The optimisation is shown to provide a natural minimum sensitivity condition, which is compared with the principle of minimum sensitivity.~\cite{Stevenson,PMS} The question of predicitive power, an interesting connection between an optimised ERG and the Polchinski RG, and further applications are discussed as well. The computation of critical exponents for the Ising model serves as an illustration for our various findings.

\section{Flows, regulators and the gap}

The modern way of implementing an ERG goes by adding a regulator term quadratic in the fields  $\sim \int d^dq \phi(-q) R_k(q^2)\phi(q)$ (for bosonic fields) to the action.~\cite{CW,Morris:1994qb} The operator $R_k(q^2)$, sometimes referred to as the regulator scheme (RS), induces a scale dependence, which, when written for the scale-dependent effective action $\Ga_k$, results in the flow equation
\beq \label{general}
\0{\partial}{{\partial t}}\Ga_k[\phi]
=\012\Tr\left\{
\left(\0{\delta^2\Gamma_k[\phi]}{\delta\phi(q)\delta\phi(-q)}+R_k(q^2) \right)
^{-1}\0{\partial R_k}{{\partial t}}\right\}. 
\eeq
Here,  $\phi$  denotes bosonic fields and  $t = \ln k$ the logarithmic scale parameter.  The right hand side of \eq{general}  contains the full inverse propagator and the trace denotes a sum over all indices and integration over all momenta. 

The RG flow is specified through the operator $R_k(q^2)$, which can be chosen at will within some basic restrictions. First of all, it is required that $R_k(q^2\to 0)>0$. This ensures that the effective propagator at vanishing field remains finite in the infrared limit $q^2\to 0$, and no infrared divergences are encountered in the presence of massless modes. The second requirement is the vanishing of $R_k$ in the infrared, $R_k(q^2)\to 0$ for $k\to 0$. This guarantees that the scale-dependent effective action $\Gamma_k$ reduces to the quantum effective action $\Ga=\lim_{k\to 0}\Ga_k$. The third condition to be met is that $R_k(q^2)$ diverges in the UV limit $k\to\Lambda$. This way it is ensured that $\Ga_k$ approaches the microscopic action $S=\lim_{k\to \Lambda}\Ga_k$ in the ultraviolet limit $k\to \Lambda$. These conditions guarantee that the flow \eq{general} interpolates between the classical and the quantum effective action. 

The main ingredient of the flow equation is the full regularised propagator. Its inverse differs from the `free' one in an important aspect: it displays a gap as a function of momenta and remains strictly positive for $k>0$,
\beq
\min_{q^2\ge 0}
\left(\left.\0{\delta^2\Gamma_k[\phi]}{\delta\phi(q)\delta\phi(-q)}
\right|_{\phi=0}+R_k(q^2) 
\right)
= C k^2 \,.
\eeq 
The existence of the gap $C>0$ is an immediate consequence of $R_k$ being an infrared regulator as defined above. Its size depends on the regulator function. It can be made arbitrarily small, which corresponds to the removal of the regulator, but not arbitrarily large, $C<\infty$. This observation leads naturally to an extremisation condition which is the main topic of this contribution. In terms of the dimensionless inverse propagator at vanishing field $P^2$, the gap is given by 
\beq\label{Cb}
C=\min_{y\ge 0} P^2(y) \ .
\eeq
An explicit expression for $P^2$ is $P^2(y)=y[1+r(y)]$, where we have written the regulator in terms of a dimensionless function $r(y)$ as $R_k(q^2)=q^2\, r(y)$ and $y=q^2/k^2$. For simplicity, we also assumed a standard kinetic term. Higher order effects like wave-function renormalisations can be considered as well.~\cite{Litim:2000ci,Exponents}

\section{Optimisation}

A simple extremisation condition based only on $P^2(y)$ follows from requiring the gap to be maximal with respect to the regulator scheme,
\beq\label{opt}
C_{\rm opt}=\max_{\rm RS}\left(\min_{y\ge 0} P^2(y)\right) \ .
\eeq
We denote those regulators as `optimal' for which the maximum is attained. The criterion \eq{opt} is based only on the effective propagator at vanishing field which renders the optimisation condition universally applicable. No reference is made to a specific model or theory. Notice also that any regulator function $R_k$ can be characterised by a countably infinite set of parameters (because $R_k(q^2)$ is at least square-integrable), of which only one is fixed by the optimisation criterion \eq{opt}.
 
In ref.~\cite{Litim:2000ci}, we have provided a number of physical interpretations of the criterion \eq{opt}, linked to the convergence of amplitude expansions, the convergence of the derivative expansion, the approach to convexity and the pole structure of threshold functions. The perhaps simplest physical explanation is given as an expansion of the flow equation in inverse powers of $P^2(y)$. Such expansions {\it always} exist, simply because $P^2$ displays a gap. The corresponding expansion coefficients~\cite{Litim96} are moments of a scheme-dependent kernel $K$ w.r.t.~powers of $1/P$,
\beq\label{ak}
a_n=\int_0^\infty dy\, K[r]\, P^{-n}(y)\ .
\eeq 
Explicit examples for $K$, which depends on the specific physical quantity studied, are given in ref.~\cite{Litim:2000ci}. Here, we only need to know that the kernel $K$ for general expansion coefficients is finite, peaked, and suppressed for sufficiently large momenta $q^2/k^2$. This is a direct consequence of the constraints for $R_k$ and the structure of the flow. 

Let us consider the large-$n$ behaviour of the coefficients \eq{ak}. The factor $P^{-n}$ strongly suppresses $a_n$ in the limit $n\to\infty$ because $P^2$ itself displays a gap and diverges for large momenta. The sole contribution to the integrand will then come from the minimum of $P^2$ where the integrand is the least suppressed. Taking into account that the kernel $K$ is well-behaved, we can conclude that the size of the pole \eq{Cb} determines the size of the expansion coefficient
\beq\label{an-large}
a_n\sim C^{-n/2}
\eeq 
for sufficiently large $n$, apart from a $K$-dependent (but $n$-independent) numerical prefactor. This brings us directly to the radius of convergence for amplitude expansions, defined as the ratio $a_n/a_{n+2}$ of two successive expansion coefficients in the limit $n\to\infty$. Making use of \eq{an-large} it is found that the radius is given by the size of the gap \eq{Cb}, independent of $K$. Hence, the condition for maximising the radius of convergence coincides with the optimisation condition \eq{opt}. 
 
\section{Optimisation vs. scheme dependence}

The optimisation condition can be seen as a {\it natural} minimum sensitivity condition, linked to the spurious scheme dependence of approximate solutions to the flow equation. This statement deserves some explanation. Let us consider the scheme dependence of a physical observable $O_{\rm phys}$. Without loss of generality, we may assume that the effective action has been parametrised by a set of `couplings' $\lambda_m$. Also, the scheme is parametrised by sets of numbers $a_n$. Solving the flow equation for a given set of initial values gives the couplings $\lambda_m$ as functions of the initial values and of the parameters $a_n$. Physical observables are functions of the couplings $\lambda_m$. Introducing the notation `(RS)' for a parametrisation of different classes of regulators, the spurious scheme dependence of $O_{\rm phys}$ is
\beq\label{dO}
\0{dO_{\rm phys}}{d({\rm RS})}=\sum_{\lambda_m,a_n}
\0{dO_{\rm phys}}{d\lambda_m}\,
\0{d\lambda_m}{da_n}\,
\0{da_n}{d({\rm RS})}\ .
\eeq
The factors ${dO_{\rm phys}}/{d\lambda_m}$ and ${d\lambda_m}/{da_n}$ are in general non-vanishing and solely determined by the physical problem and the chosen parametrisation of the effective action. The factors $da_n/d({\rm RS})$ encode the essential scheme dependence. In general all terms in \eq{dO} are non-vanishing and scheme-dependent. For the full solution (no approximations) the sum vanishes for all (RS). For approximate solutions, \eq{dO} does not vanish automatically for all (RS). 
Let us consider the factors $da_n/d({\rm RS})$ in more detail. Within an amplitude expansion we can use the coefficients \eq{ak} (with possibly different kernels) to parametrise the regulator. Their RS dependence is given by
\beq \label{dak}
\0{d a_n}{d ({\rm RS})} = 
\int _0^\infty dxdy
\left[\0{\delta K[r](x)}{y\delta r(y)}
      -\0n2\delta(x-y)\0{K[r](y)}{y[1+r(y)]} \right]\,
P^{-n}(y)\,\0{dP^2(y)}{d({\rm RS})}\ .
\eeq
In general, the factor in square brackets does not vanish. However, for $n$ sufficiently large the integrand is strongly peaked due to the factor $P^{-n}$ which gives the main contribution to the integrand at its maximum -- which is the minimum of $P^2(y)$. In this case the sole contribution to the integral comes  from the region around the minimum of $P^2(y)$. At this minimum we can replace $dP^2(y)/d({\rm RS})$ by $dC/d({\rm RS})$. Apart from a (finite) numerical prefactor, we obtain for \eq{dak}, consistent with \eq{an-large}, 
\beq \label{dak2}
\0{d a_n}{d({\rm RS})} \sim \0{dC}{d({\rm RS})}\,.
\eeq
The r.h.s.~of \eq{dak2} vanishes for optimised regulators, due to \eq{opt}. Hence, we have just shown that the optimisation removes, for a given class of regulators, the essential scheme dependence due to \eq{dak}. Therefore it provides a natural minimum sensitivity condition, regardless of the truncation or the theory or the observable considered. 

\section{Optimisation vs. principal of minimum sensitivity}

Let us contrast the optimisation condition with the well-known principal of minimum sensitivity (PMS)~\cite{Stevenson,PMS}, originally invented in the context of perturbative QCD. Given a class of regulators, the PMS condition singles out those schemes which make \eq{dO} vanish for the truncated case. This way, the spurious dependence on unphysical parameters is removed. The straightforward use of a PMS condition for flow equations has, however, a number of shortcomings: a solution to the PMS condition cannot be guaranteed; it may not be unique; it generically is non-universal, depending on the specific theory, on the specific observable, on the truncation and on the class of regulator function used for the extremisation.

To be more explicit, consider the Ising universality class, a $N=1$-component real scalar field theory in $3d$ at the Wilson-Fisher fixed point. The critical exponent $\nu_{\rm ERG}$ is the inverse of the single positive eigenvalue of the stability matrix at criticality. To leading order in a derivative expansion it depends on the scheme. We make use of the results of ref.~\cite{LPS}, where the authors solved the PMS condition numerically within a polynomial expansion of the scaling potential for three classes of regulators parametrised as $r_{\rm exp}=1/[2^{y^b} -1] $, $r_{\rm power}=y^{-b}$ and $r_{\rm mix}=\exp \left[-b(y^{1/2}-y^{-1/2})\right]$, all for $b\ge 1$. One solution $b_{\rm PMS}$ for each regulator, with a very weak dependence on the truncation, has been reported explicitly in ref.~\cite{LPS}. A second solution exists for $r_{\rm exp}$ and $r_{\rm power}$ in the limit $b\to\infty$, given by the sharp cut-off regulator.~\cite{PMS} In order to facilitate a comparison with the corresponding optimised regulator, we convert the PMS parameters $b_{\rm PMS}$ into gaps $C_{\rm PMS}$, using \eq{Cb}. In Tab.~1, the ratios $C_{\rm PMS}/C_{\rm opt}$ are given for all different solutions to the PMS condition. \\
\begin{center}
\begin{minipage}{.42\hsize}
{\label{Tab1}
\small 
Tab.~1: Optimisation vs.~PMS for specific regulators: two PMS solutions for $r_{\rm exp}, r_{\rm power} $ and one for $r_{\rm mix}$ (see text).}
\end{minipage}
\hskip.1cm
\begin{tabular}{l|c|c|c}
{\small Regulator}& $r_{\rm exp} $ & $r_{\rm power} $& $r_{\rm mix}$
\\[1ex] 
\hline
&&&\\[-2ex]
$\small C_{\rm PMS}/C_{\rm opt}$
&$0.999\,,  \  \s012$
&$0.963\,,  \  \s012$
&$0.984$
\\ 
\end{tabular}
\end{center}
\vskip.4cm
There are a few lessons to be learnt from Tab.~1. Solutions to the PMS condition indeed depend on the chosen set of regulator functions and, though weakly, on the truncation. Furthermore, it can be shown that the PMS condition is solved by the sharp cut-off for any theory, any observable and any truncation.~\cite{PMS} Hence, the PMS condition by itself is not sufficient to single-out a unique solution and additional criteria (like convergence properties~\cite{LPS}) are required. 

In turn, the optimisation condition \eq{opt} clearly discards the sharp cut-off simply because $C_{\rm sharp}/C_{\rm opt}=\s012\neq 1$. More importantly, we notice that one solution of the PMS condition is always very close to the corresponding solution obtained from optimising the propagator gap! This implies that the corresponding critical exponents obey $\nu_{\rm PMS}/\nu_{\rm opt}\approx 1$, a direct consequence of the vanishing of \eq{dak2} for optimised regulators. This shows explicitly that the optimisation provides a {natural} minimum sensitivity condition, which, for the above reasons, has more predictive power than the standard PMS condition. For a detailed discussion, see ref.~\cite{PMS}.

\section{Optimisation vs.~predictive power and outlook}

In order to address the issue of predictive power, we ressort to the example of the previous section. Let us elaborate on the conjecture alluded to in the beginning, namely that the values $\nu_{\rm opt}$ are in the vicinity of the physical value $\nu_{\rm phys}$. More generally, the conjecture applies for generic observables. It is substantiated by our above analysis, which adds two important pieces of information: $(i)$ the critical exponent $\nu_{\rm ERG}$, a scheme-dependent number, is {\it bounded} as a function of the scheme~\cite{Exponents}; $(ii)$ the boundary value $\nu_{\rm opt}$ is attained for $r_{\rm opt}=(\s01y -1)\Theta(1-y)$ (see ref.~\cite{Exponents} for the derivation of $r_{\rm opt}$). Explicitly, we find that $\nu_{\rm ERG}\ge \nu_{\rm opt}$ with $\nu_{\rm opt}=0.64956\cdots$, while the physical value is given by $\nu_{\rm phys}\approx 0.625$ (Ising universality class). Here, the optimised  value is indeed the closest to the physical one.

These particular results are interesting for another reason. We have computed the first few eigenvalues at criticality in the optimised case, including the critical exponent $\nu_{\rm opt}$ for all $N$.~\cite{Exponents} Our results agree, to all published digits, with the corresponding ones from the Polchinski RG~\cite{Polchinski,Comellas}. This is a very surprising result for at least two reasons. First of all, the ERG and the Polchinski RG, linked by a Legendre transform, have inequivalent derivative expansions. Secondly, the Polchinski RG is scheme-independent to leading order in the derivative expansion~\cite{Ball:1995ji} while the ERG isn't. Our findings point towards a more subtle picture: the ERG contains a redundant scheme freedom, which, when removed, makes the results of the two RGs equivalent in the present example. 

The present analysis closes a gap in the ERG formalism by providing an understanding of the spurious scheme dependence and the related convergence of approximate solutions. This control of the flow is mandatory for reliable physical predictions. Extensions of these considerations to the case of fermions or gauge fields are straightforward.~\cite{Litim:2000ci} For gauge theories, modified Ward or BRST identities~\cite{gaugefields,Freire:2000bq,axial} ensure the gauge invariance of physical Green functions, and the optimisation criterion is compatible with such an additional constraint. Also, the wave function renormalisation can be taken into account in the usual manner without changing the optimisation condition.~\cite{Litim:2000ci,Exponents} This optimisation works also for field theories at finite temperature within the imaginary or the real-time formalism.~\cite{Litim98} Finally, it would be interesting to see how these ideas apply to Hamiltonian flows~\cite{Wegner}.

\section*{Acknowledgements}

It is a pleasure to thank the organisers for a stimulating conference. Financial support from the University of Heidelberg is gratefully acknowledged.

\section*{References}

%************************************************************************|
% Bibliography
%************************************************************************|

\def\PRA#1#2#3#4#5{ #1,\,\,{\it }\,Phys.\,Rev.\,{\bf A#3}\,(19#4)\,#5}
\def\PRB#1#2#3#4#5{#1,{\it }\,Phys.\,Rev.\,{\bf B#3}\,(19#4)\,#5}
\def\PRL#1#2#3#4#5{#1,\,\,{\it }\,Phys.\,Rev.\,Lett.\,{\bf #3} (19#4) #5}
\def\PRC#1#2#3#4#5{#1,\,\,{\it }\,Phys.\,Rev.\,{\bf C#3}\,(19#4)\,#5}
\def\PRD#1#2#3#4#5{#1,{\it }\,Phys.\,Rev.\,{\bf D#3}\,(19#4)\,#5}
\def\PRE#1#2#3#4#5{#1,\,\,{\it }\,Phys.\,Rev.\,{\bf E\,#3}\,(19#4)\,#5}
\def\PRep#1#2#3#4#5{#1,{\it }\,Phys.\,Rep.\,{\bf  #3}\,(19#4)\,#5}
\def\NPB#1#2#3#4#5{#1,{\it }\,Nucl.\,Phys.\,{\bf B#3}\,(19#4)\,#5}
\def\PLB#1#2#3#4#5{#1,{\it }\,Phys.\,Lett.\,{\bf B#3}\,(19#4)\,#5}
\def\JPA#1#2#3#4#5{#1,\,\,{\it }\,J.\,Phys.\,{\bf A#3}\,(19#4)\,#5}
\def\JPB#1#2#3#4#5{#1,\,\,{\it }\,J.\,Phys.\,{\bf B#3}\,(19#4)\,#5}
\def\JPC#1#2#3#4#5{#1,\,\,{\it }\,J.\,Phys.\,{\bf C#3}\,(19#4)\,#5}
\def\ZPC#1#2#3#4#5{#1,{\it }\,Z.\,Phys.\,{\bf C#3}\,(19#4)\,#5}
\def\MPLA#1#2#3#4#5{#1,{\it }\,Mod.\,Phys.\,Lett.\,{\bf A#3}\,(19#4)\,#5}
\def\and#1#2#3{{\bf #1}\,(19#2)\,#3}

\end{document}